\documentclass[multi]{cambridge7A}

\usepackage[UKenglish]{babel}
\usepackage[longnamesfirst,sectionbib]{natbib}
\usepackage{chapterbib}
\usepackage{amsmath,amssymb,amsthm,amsfonts}
\usepackage{enumerate,url}
\usepackage{graphicx}      %Allows one to include graphics in document
\usepackage{rotating}	%rotate tables into landscape format

\setcitestyle{numbers,square,comma}% alters natbib behaviour

\allowdisplaybreaks[1]

\newcommand{\E}{\operatorname{E}}
\newcommand{\prob}{\mathbf{P} }   %probability

\providecommand*\Index[1]{#1\index{#1}}
\providecommand*\undex[1]{} % abandoned tag
\begin{document}
\alphafootnotes
\author[Peter Donnelly and Stephen Leslie]{Peter Donnelly\footnotemark\ and
  Stephen Leslie\footnotemark }
\chapter[The coalescent and its descendants]{The coalescent and its
  descendants}
\footnotetext[1]{Wellcome Trust Centre for Human Genetics, University of
  Oxford, Roosevelt Drive, Oxford OX3 7BN, and Department of Statistics, 1
  South Parks Road, Oxford OX1 3TG; donnelly@stats.ox.ac.uk}
\footnotetext[2]{Department of Statistics, University of Oxford, 1 South
  Parks Road, Oxford OX1 3TG; leslie@stats.ox.ac.uk}
\arabicfootnotes
\contributor{Peter Donnelly \affiliation{University of Oxford}}
\contributor{Stephen Leslie \affiliation{University of Oxford}}
\renewcommand\thesection{\arabic{section}}
\numberwithin{equation}{section}
\renewcommand\theequation{\thesection.\arabic{equation}}
\numberwithin{figure}{section}
\renewcommand\thefigure{\thesection.\arabic{figure}}
\numberwithin{table}{section}
\renewcommand\thetable{\thesection.\arabic{table}}

\begin{abstract}
The coalescent revolutionised theoretical
population genetics, simplifying, or making possible for the
first time, many analyses, proofs, and derivations, and
offering crucial insights about the way in which the structure
of data in samples from populations depends on the demographic
history of the population.  However statistical inference under the
coalescent model is extremely challenging, effectively because
no explicit expressions are available for key sampling
probabilities.  This led initially to approximation of these
probabilities by ingenious application of modern
computationally-intensive statistical methods.  A key
breakthrough occurred when Li and Stephens introduced a
different model, similar in spirit to the coalescent, for which
efficient calculations are feasible.  In turn, the Li and
Stephens model has changed statistical inference for the wealth
of data now available which documents molecular genetic
variation within populations.  We briefly review the coalescent
and associated measure-valued diffusions, describe the Li and
Stephens model, and introduce and apply a generalisation of it
for inference of population structure in the presence of
linkage disequilibrium.  
\end{abstract}

\subparagraph{AMS subject classification (MSC2010)}60J70, 62M05, 92D10

\section{Introduction}\label{introduc}

John Kingman\index{Kingman, J. F. C.!influence|(} made a number of incisive
and elegant contributions to modelling in the field of
genetics\index{mathematical genetics|(}, several of which are described
elsewhere in this volume.  But it is probably the coalescent, or `Kingman
coalescent'\index{Kingman, J. F. C.!Kingman coalescent|(} as it is often known,
which has had the greatest impact.  Several authors independently developed
related ideas around the same time
\cite{Griffiths:1980rt}\index{Griffiths, R. C.},
\cite{Tajima:1983fp}\index{Tajima, F.},
\cite{Hudson:1983yf}\index{Hudson, R. R.} but it was Kingman's
description and formulation, together with his proofs of the key robustness
results \cite{Kin1982C}, \cite{Kin1982A}, \cite{Kin1982B} which had the
greatest impact in the mathematical genetics community.

More than 25 years later the coalescent remains central to much
of population genetics, with book-level treatments of the subject
now available \cite{Tav2004}\index{Tavar\'e, S.|(},
\cite{HSW2005}\index{Hein, J.}\index{Schierup, M. H.}\index{Wiuf, C.},
\cite{Wakeley2009}\index{Wakeley, J.}.
As others have noted, population genetics as a field was theory rich and
\index{data!availability}data
poor for much of its history.  Over the last five years this has changed
beyond recognition.  The development and wide application of high-throughput
experimental techniques for assaying\index{assay} molecular
\Index{genetic variation} means that scientists are now awash with data.
Enticing as this seems, it turns out that the coalescent model cannot be fully
utilised for analysis of these data---it is simply not computationally
feasible to do so.  Instead, a closely related model, due to Li and Stephens,
has proved to be computationally tractable\index{computational complexity} and
reliable as a basis for inference for modern genetics data.  Alternative
approaches are based on approximate inference under the coalescent\index{coalescence!inference}.

Our purpose here is to give a sense of these developments, before describing
and applying an extension of the Li and Stephens
model\index{LiN@Li, N.!Li and Stephens model} to
\index{population structure!geographical population structure}populations
with geographical structure.  We do not attempt an extensive review.

The initial historical presentation is necessarily somewhat technical in
nature, and provides some explanation of the theoretical developments leading
up to the Li and Stephens model.  Readers interested in just the Li and
Stephens model and its application may begin at Section \ref{recombination} as
the presentation of this material does not heavily depend on the previous
sections.

Before doing so, one of us (PD) will indulge in a brief personal reminiscence.
John Kingman was my doctoral supervisor.  More accurately, he acted as my
supervisor for a year, before leaving Oxford\index{Oxford, University of} to
Chair the then UK
\Index{Science and Engineering Research Council} (I have always hoped the
decision to change career direction was unrelated to his supervisory
experiences).  During the year in question, Kingman wrote his three seminal
coalescent papers.  A photocopy of one of the manuscripts, in John's
immaculate handwriting, unblemished by corrections or second thoughts as to
wording, still survives in my filing cabinet.

There is a certain irony to the fact that although the
coalescent was the unifying theme of much of my academic work
for the following 20 years, it formed no part of my work under
Kingman's supervision.   John's strategy with new research
students, or at least with this one, was to direct them to the
journals in the library, note that some of the papers therein
contained unsolved problems, and suggest that he would be happy
to offer advice or suggestions if one were stuck in solving one
of these.  This was daunting, and the attempts were largely
unsuccessful.  It was only as Kingman was leaving Oxford that he
passed on copies of the coalescent
manuscripts\index{Kingman, J. F. C.!Kingman coalescent|)}, and, embarrassingly,
it was some years before I saw the connection between the coalescent
and aspects of my doctoral work on interacting particle
systems\index{interacting particle system}, then under Dominic
Welsh's\index{Welsh, D. J. A.}
supervision.\index{Kingman, J. F. C.!influence|)}

\section{The coalescent and the Fleming--Viot process}\label{models}

To set the scene, we briefly outline the context in which the coalescent\index{coalescence|(}
arises, and then describe the coalescent itself along with the corresponding
process forward in time, the so-called Fleming--Viot measure-valued
diffusion\index{diffusion!measure-valued diffusion}\index{Fleming, W. H.!Fleming--Viot diffusion}.
We aim here only to give a brief flavour of the two processes rather than a
detailed exposition.

The most basic, and oldest, models in population genetics are finite Markov
chains\index{Markov, A. A.!Markov chain} which describe the way in which the
genetic composition of the population changes over time.  In most cases, these
models are not tractable, and interest moves to their limiting behaviour as
the population size grows large, under suitable re-scalings of time.  When
examined forward in time, this leads to a family of measure-valued diffusions,
called Fleming--Viot processes.  In a complementary, and for many purposes
more powerful, approach one can instead look backwards in time, and focus on
the genealogical tree relating sampled chromosomes\index{chromosome|(}.  In the
large population limit, these (random) trees\index{random tree} converge to a
particular process called the coalescent.

We start with the simplest setting in which individuals are
\emph{haploid}\index{haploid}; that is they carry a single copy of their
genetic material which is inherited from a single parent.  Many organisms,
including humans, are \emph{diploid}\index{diploid}, with each
\Index{cell} carrying two copies of the individual's
\index{DNA (deoxyribonucleic acid)}DNA---these
copies being inherited one from each of the individual's two parents.  It
turns out that the haploid models described below also apply to diploid
organisms provided one is interested in modelling the evolution of small
contiguous segments of DNA---for many purposes we can ignore the fact that in
diploid organisms these small segments of DNA occur in pairs in individuals,
and instead model the population of individual chromosomes, or more precisely
of small segments of them taken from the same genomic region, in each
individual. In what follows, we will respect this particular perspective and
refer to the haploid `individuals' in the population we are modelling as
`chromosomes'.

One simple discrete model for population demography\index{demography|(} is the
Wright--Fisher model\index{Wright, S.!Wright--Fisher model|(}.  Consider a
population of fixed size $N$ chromosomes which evolves in discrete
generations.  The random mechanism for forming the next generation is as
follows: each chromosome in the next generation chooses a chromosome in the
current generation (uniformly at random) and copies it, with the choices made
by different chromosomes being independent.  An equivalent description is that
each chromosome in the current generation gives rise to a random number of
copies in the next generation, with the joint distribution of these offspring
numbers being symmetric multinomial.

In addition to modelling the demography of a population, a population genetics
model needs to say something about the genetic types\index{type|(} carried by the
chromosomes in the population, and the way in which these change
(probabilistically) when being copied from parental to offspring chromosomes.
Formally, this involves specifying a set, $E$, of possible types (usually, if
unimaginatively, called the \emph{type space})\index{type!type space}, and a
matrix of transition probabilities $\Gamma$ whose $i$, $j$th entry,
$\gamma_{ij}$, specifies for each $i$, $j \in E$, the probability that an
offspring chromosome will be of type $j$ when the parental chromosome is of
type $i$.  The generality has real advantages: different choices of type space
$E$ can be used for modelling different kinds of genetic information. In most
genetic contexts, offspring are extremely likely to have the same type as
their parent, with changes to this type, referred to
as \emph{mutations}\index{mutation|(}, being extremely rare.

Under an assumption of genetic neutrality\index{mathematical genetics!neutrality|(}, all variants in
a population are equally fit and are thus equally likely to be transmitted.
This assumption allows a crucial simplification: the random process describing
demography is independent of the genetic types carried by the individuals in
the population.  In this case, one can first generate the demography of the
population using, say, the Wright--Fisher model, and then independently
superimpose the genetic type for each chromosome, and the details of the
(stochastic) mutation process which may change types.  The separation of
demography from genetic types lies at the heart of the simplification offered
by the coalescent: the coalescent models the parts of the demography relevant
to the population at the current time; information about genetic
types can be added independently.  The extent to which the
neutrality assumption applies is rather controversial in general, and for
humans in particular, but it seems likely that it provides a reasonable
description for many parts of the \Index{genome}.

The Wright--Fisher model may also be extended to allow for more realistic
demographic effects, including variation in population size, and geographical
spatial structure in the 
\index{population structure!geographical population structure}population (so
that offspring
chromosomes\index{chromosome|)} are more likely to be located near to their
parents).  We will not describe these here.  Somewhat surprisingly, it
transpires that the simple model described above, (constant population size,
random mating, and neutrality---the so-called `standard
neutral'\index{mathematical genetics!standard neutral model} model), or rather its large
population limit, captures many of the important features of the evolution of
human and other populations. There is an aphorism in
\index{statistical inference}statistics that
``all models are false, but some are useful''.  The standard neutral model has
proved to be extremely useful.\index{mathematical genetics!neutrality|)}

In a Wright--Fisher, or any other, model, we could describe the genetic
composition of the population at any point in time by giving a list of the
genetic types currently present, and the proportion of the population
currently of each type.  Such a description corresponds to giving a
probability measure on the set $E$ of possible types.  It is sometimes helpful
to think of this measure as the distribution of the type of an individual
picked at random from the population. Note that summarising the population
composition in this way at a particular time point involves an assumption
of exchangeability\index{exchangeability|(} across individuals: it is only
the types present,
and the numbers of individuals of each type, in a particular population which
matter, with information about precisely which individuals carry particular
types not being relevant. In this framework, when we add details of the
mutation process to the Wright--Fisher model, by specifying $E$ and $\Gamma$,
we obtain a discrete time (probability-) measure-valued Markov
process\index{Markov, A. A.!Markov process}.  As $N$ becomes large a suitable
rescaling of the process converges to a diffusion limit: time is measured in
units of $N$ generations, and mutation probabilities, the off-diagonal entries
of the matrix $\Gamma$ above, are scaled as $N^{-1}$.  For general genetic
systems, the limit is naturally formulated as a measure-valued process, called
the Fleming--Viot diffusion\index{Fleming, W. H.!Fleming--Viot diffusion|(}.
The classical so-called Wright--Fisher
diffusion\index{Wright, S.!Wright--Fisher diffusion} is a one-dimensional
diffusion on $[0,1]$ which arises when there are only two possible genetic
types and one tracks the population frequency of one of the types. This is a
special case of the Fleming--Viot diffusion, in which we can identify the
value of the classical diffusion, $p \in [0,1]$, with a probability measure on
a set with just two elements. The beauty of the more general, measure-valued,
formulation is that it allows much more complicated genetic types, which could
track
\index{DNA (deoxyribonucleic acid)}DNA sequences, or more exotically even
keep track of the time since particular mutations arose in the population.

The Fleming--Viot process can thus be thought of as an
approximation to a large population evolving according to the
Wright--Fisher model. As noted, for the Wright--Fisher model,
time is measured in units of $N$ generations in this
approximation (and the approximation applies when mutation
probabilities are of order $N^{-1}$). In fact the Fleming--Viot
process arises as the limit of a wide range of demographic
models (and we refer to such models as being within the
\Index{domain of attraction} of the Fleming--Viot process), although the
appropriate time scaling can differ between models. (See, for
example, \cite{EtK1993}\index{Ethier, S. N.}\index{Kurtz, T. G.}.) For
background, including explicit formulations of the claims made above, see
\cite{DoK1996}, \cite{DoK1999} \cite{EtK1986}, \cite{EtK1993},
\cite{Ewe2004}\index{Ewens, W. J.}.

Donnelly and Kurtz \cite{DoK1996}, \cite{DoK1999} give a
discrete construction of the\break Fleming--Viot process.  As a
consequence, the process can actually be thought of as
describing the evolution of a hypothetically infinite
population, and it explicitly includes the demography of that
population.  Exchangeability figured prominently in Kingman's
work in genetics.  It provides a linking thread here: the
Donnelly--Kurtz construction embeds population models for each
finite population size $N$ in an infinite exchangeable
sequence.  The value taken by the Fleming--Viot
diffusion\index{Fleming, W. H.!Fleming--Viot diffusion|)} at a
particular time point is just the de Finetti\index{Finetti, B. de}
representing measure for the infinite exchangeable\index{exchangeability|)}
sequence. Given the value of the measure, the types of individuals in the
population are independent and identically distributed according to that
measure.

The coalescent arises by looking backwards in time.  Consider again the
discrete Wright--Fisher model.  If we consider two different
chromosomes\index{chromosome|(} in the current generation, they will share an
ancestor
in the previous generation with probability $1/N$.  If not, they
retain distinct ancestries, and will share an ancestor in the
 generation before that with probability $1/N$.  The number of
generations until they share an ancestor is thus geometrically
distributed with success probability $1/N$ and mean $N$.  In the
limit for large $N$, with time measured in units of $N$
generations, this geometric random variable converges to an
exponential random variable with mean 1.

More generally, if we consider $k$ chromosomes, then for fixed $k$
and large $N$, they will descend from $k$ distinct ancestors in
the previous generation with probability
\begin{equation} \nonumber
1 - \binom k2\frac{1}{N} + O(N^{-2}).
\end{equation}
Exactly two will share a common ancestor in the previous
generation with probability $\binom k2\frac{1}{N} + O(N^{-2})$,
and more than a single pair will share a common ancestor with
probability $O(N^{-2})$.  In the limit as $N \to \infty$, with
time measured in units of $N$ generations, the time until any of
the $k$ share an ancestor will be exponentially distributed with
mean $1/{\binom k2}$, after which time a randomly chosen pair
of chromosomes\index{chromosome|)} will share an ancestor.

Thus, in the large population limit, with time measured in
units of $N$ generations, the genealogical history of a sample
of size $n$ may be described by a random binary tree\index{random tree}. The
tree initially has $n$ branches, for a period of time $T_n$,
after which a pair of branches (chosen uniformly at random,
independently of all other events) will join, or coalesce.
More generally, the times $T_k$, $k = n$, $n-1$, \ldots, 2 for
which the tree has $k$ branches are independent exponential
random variables with
\begin{equation}\nonumber
\E(T_k)={\binom k2}^{-1},
\end{equation}
after which a pair of branches (chosen uniformly at random independently of
all other events) will join, or coalesce.  The resulting random tree is called
the $n$-coalescent, or often just the coalescent.  Note that we have described
the coalescent as a random tree.  Kingman's original papers elegantly
formulated the $n$-coalescent as a stochastic process on the set of
equivalence relations\index{equivalence class/relation} on
$\{1,2, \ldots,n\}$.  The
two formulations are equivalent.  We view the tree description as more
intuitive\index{coalescent tree}.

In a natural sense the tree describes the important part of the
genealogical history of the sample, in terms of their genetic
composition.  It captures their shared ancestry, due to the
demographic process.  As noted above, a key observation is that
in neutral models the distribution of this ancestry is
independent of the genetic types which happen to be carried by
the individuals in the population. Probabilistically, one can
thus sample the coalescent tree and then superimpose genetic
types.  For example, at stationarity, first choose a type for
the most recent common ancestor of the population (the type at
the root of the coalescent tree) according to the stationary
distribution of the mutation process, and then track types
forward through the tree from the common ancestor, where they
will possibly be changed by mutation\index{mutation|)}.

The preceding recipe gives a simple means of simulating the
genetic types of a sample of size $n$ from the population.
Note that this is an early example of what has more recently
come to be termed `exact simulation'\index{simulation!exact/perfect simulation}: a
finite amount of
simulation producing a sample with the exact distribution given
by the stationary distribution of a Markov
process\index{Markov, A. A.!Markov process}.  In
addition, it is much more computationally efficient than
simulating the entire population forward in time for a long
period and then taking a sample from it.  Finally, it reveals
the complex structure of the distribution of genetics models at
stationarity---the types of each of the sampled chromosomes\index{chromosome|(}
are (positively) correlated, exactly because of their shared
\Index{ancestral history}.

We motivated the coalescent from the
Wright--Fisher model\index{Wright, S.!Wright--Fisher model|)}, but the
same limiting genealogical tree arises for any of the large class
of demographic models in the \Index{domain of attraction} of the
Fleming--Viot diffusion\index{Fleming, W. H.!Fleming--Viot diffusion}. See
Kingman \cite{Kin1982C}\index{Kingman, J. F. C.} for an elegant formulation
and proof of this kind of robustness result. Moreover, the ways in which the
tree shape changes under different demographic scenarios (e.g.\ changes in
population size or geographical
\index{population structure!geographical population structure}population
structure) is well
understood \cite{Wakeley2009}\index{Wakeley, J.},
\cite{HSW2005}\index{Hein, J.}\index{Schierup, M. H.}\index{Wiuf, C.}.

The discrete construction of the Fleming--Viot process described
above actually embeds the coalescent and the forward diffusion in
the same framework, so that one can think of the coalescent as
describing the \Index{genealogy} of a sample from the diffusion.

There is even a natural limit, as $n \to \infty$, of the
$n$-coalescents\index{coalescence!n coalescents@$n$-coalescents},
introduced and studied by Kingman \cite{Kin1982B}.  This
can be thought of as the limit of the genealogy of the whole
population, or as the genealogy of the infinite population
described by the Fleming--Viot process (though this perspective
was not available when Kingman introduced the process). The
analysis underlying the relevant limiting results for this
population-genealogical process is much more technical than
that outlined above for the fixed-sample-size
case \cite{donnelly1991weak}, \cite{donnelly1992weak}\index{Joyce, P.},
\cite{DoK1996}\index{Kurtz, T. G.}.   It is easiest to describe this tree
from the
root, representing the common ancestor of the population,
forward to the tips, each of which represents an individual
alive at the reference time. The tree has $k$ branches for a
random period of time $T_k$, after which a branch, chosen
uniformly at random, independently for each $k$, splits to form
two branches. The times $T_k$, $k=2$, 3, \ldots, are
independent exponential random variables, and independent of
the topology of the tree, with \[ \E(T_k)={\binom k2}^{-1}. \]
Write \begin{equation}\nonumber \label{T} T = \sum_{k=1}^\infty
T_k \end{equation} for the total depth of the tree, or
equivalently for the time back until the population first has a
common ancestor.  Note that $T$ is a.s.\ finite.  In fact
$\E(T)=2$.

To date, we have focussed on models for a small segment of
\index{DNA (deoxyribonucleic acid)}DNA.  For
larger segments, in \Index{diploid} populations, one has to allow for the
process of \emph{recombination}\index{recombination|(}.  Consider a particular
human chromosome inherited from a parent.  The parent will have two (slightly
different) copies of this chromosome.
Think of the process which produces the chromosome to be passed on to the
offspring as starting on one of the two chromosomes in the parent and
copying from it along the chromosome.  Occasionally, and for our
purposes randomly, the copying\index{copying} process will `cross over' to the
other chromosome in the parent, and then copy from that,
perhaps later jumping back and copying from the original
chromosome, and so on.  The chromosome passed on to the offspring will thus
be made up as a mosaic of the two chromosomes in the parent.
The crossings over are referred to as recombination events.  In practice, these
recombination events are relatively rare along the chromosome: for example in
humans, there will typically be only a few recombination
events per chromosome.

The formulation described above can be extended to allow for recombination.  In
the coalescent framework, the consequence is that in going backwards in time,
different parts of a chromosome may be inherited from different chromosomes\index{chromosome|)} in
the previous generation.  One way of conceptualising this is to imagine each
position in the DNA as having its own coalescent tree, tracing the ancestry of
the DNA in that position.  This coalescent tree, marginally, will have the same
distribution as the coalescent. As one
moves along the DNA sequence, these trees for different positions
are highly positively correlated.  In fact, two neighbouring
positions will have the same tree iff there is no recombination
event between those positions, on a lineage leading to the current sample, since their joint most recent common
ancestor.  If there is
such a recombination, the trees for the two positions will be
identical back to that point, but (in general) different before
it.  The correlation structure between the trees for different
positions is complex.  For example, when regarded as a process on
trees as one moves along the sequence, it is not Markov\index{Markov, A. A.}.
Nonetheless it
is straightforward to simulate from the relevant joint
distribution of trees, and hence of sampled sequences.  The trees
for each position can be embedded in a more general probabilistic
object (this time a graph rather than a tree) called the ancestral
recombination 
graph\index{graph!ancestral recombination graph}\index{recombination|)}
\cite{GrM1996}\index{Griffiths, R. C.|(}\index{Marjoram, P.},
\cite{griffiths1997arg}.

\section{Inference under the coalescent}\label{disease}\index{coalescence!inference|(}

The coalescent has revolutionised the way we think about and analyse population
genetics models, and changed the way we simulate from these models.  There are
several important reasons for this. One is the separation, for neutral models, of
demography from the effects of mutation\index{mutation|(}. This means that many of the properties of samples taken from
genetics models follow from properties of the coalescent tree.  A second reason
is that the coalescent has a simple, indeed beautiful, structure, which is
amenable to calculation.  Most questions of interest in population genetics can
be rephrased in terms of the coalescent, and the coalescent is a fundamentally
simpler process than the traditional forwards-in-time models.

The body of work outlined above has an applied \Index{probability} flavour; some of
it more applied (for example solving genetics questions of interest), and
some more pure (for example the links with measure-valued
diffusions\index{diffusion!measure-valued diffusion}).  Historically, much of it occurred in the 10--15
years after Kingman's coalescent papers, but even today
`coalescent methods' as they have become known in population
genetics, are central to the analysis of genetics models.

If an applied probability perspective prevailed over the first 10--15 years of
the coalescent's existence, the last 10--15 years have seen a parallel
interest in statistical questions.  Since the early 1990s there has been a
steady growth in
\index{data!availability}data documenting molecular \Index{genetic variation}
in samples taken from real populations.  Over recent years this has become a
deluge, especially for humans.  Instead of trying to study probabilistic
properties of the coalescent, the statistical perspective assumes that some
data come from a coalescent model, and asks how to do
\Index{statistical inference}
for parameters in the model, or comparisons between models (for example
arising from different demographic\index{demography|)} histories for the
population).

There have been two broad approaches.  One has been to attempt to use all the
information in the data, by basing inference (in either a frequentist or
\index{Bayes, T.!Bayesian inference}Bayesian framework) on the likelihood
under the model: the probability,
regarded as a function of parameters of interest, of observing the
configuration actually observed in the sample.  This is the thread we will
follow below.
\index{full likelihood inference@full-likelihood inference|(}Full-likelihood
inference under the coalescent turns out to be a difficult problem.  A second
approach has been to summarise the information in the data via a small set of
summary statistics, and then to base inference on these statistics.  One
particular, Bayesian, version of this approach has come to be called
\index{approximate Bayesian computation (ABC)}\emph{approximate Bayesian
computation} (ABC): one approximates the full posterior distribution of
parameters of interest conditional on the data by their conditional
distribution given just the summary statistics
\cite{beaumont2002approximate}\index{Beaumont, M. A.}\index{Zhang,
W.}\index{Balding, D. J.}.

Full-likelihood inference under the coalescent is not straightforward, for a
simple reason.  Although the coalescent enjoys many nice properties, and lends
itself to many calculations, no explicit expressions are available for the
required likelihoods.  There is one exception to this, namely settings in
which the mutation probabilities, $\gamma_{ij}$, that a
chromosome\index{chromosome|(} is of type $j$ when its parent is of type $i$,
depend only on $j$.  This so-called
\index{mutation!parent-independent mutation|(}\emph{parent-independent mutation} is
unrealistic for most modern data,
notwithstanding the fact that any
\index{allele!two-allele model}two-allele model (that is, when the
\index{type!type space}type space $E$ consists of only two elements) can be
written in this
form. For parent-independent mutation models, the likelihood is multinomial.

In the absence of explicit expressions for the likelihood,
indirect approaches, typically relying on sophisticated
computational methods,\break were developed. 
\index{Griffiths, R. C.|)}\index{Tavar\'e, S.|)}Griffiths and Tavar\'{e}
were the pioneers \cite{griffiths1994ancestral}, \cite{griffiths1994sampling},
\cite{griffiths1994simulating}, \cite{griffiths1999ages}. They devised an
ingenious computational approach whereby the likelihood was expressed as a
functional of a
\index{Markov, A. A.!Markov chain}Markov chain arising from systems of
recursive equations
for probabilities of interest.
\index{Felsenstein, J.}\index{Kuhner, M. K.}\index{Yamato, J.}\index{Beerli, P.}Felsenstein \cite{felsenstein1999likelihoods}
later showed the
\index{Griffiths, R. C.!Griffiths--Tavar\'e approach (GT)|(}Griffiths--Tavar\'{e}
(GT) approach to be a particular implementation of
\index{sampling!importance sampling|(}importance sampling.  In
contrast to the GT approach, Felsenstein and colleagues developed
\index{Markov, A. A.!Markov chain Monte Carlo (MCMC)}Markov chain
Monte Carlo (MCMC) methods for evaluating coalescent likelihoods.  These were
not without challenges: the space which MCMC methods explored was effectively
that of
\index{coalescent tree}coalescent trees, and thus extremely high-dimensional,
and assessment of mixing and convergence could be fraught.  As
subsequent authors pointed out, failure of the Markov chains to mix properly
resulted in poor approximations to the likelihood
\cite{FeD2002}\index{Fearnhead, P.}.

Donnelly and Stephens \cite{StD2000}\index{Stephens, M.} adopted an importance
sampling ap-\break proach.  They reasoned that if the GT approach was
implicitly doing importance sampling, under a particular
\index{proposal distribution|(}proposal distribution which arose automatically,
then it might be possible to improve performance by explicitly
choosing the proposal distribution.  In particular, they noted
that the optimal proposal distribution was closely related to
a particular conditional probability under the coalescent,
namely the probability that an additional, $n+1$st sampled
chromosome will have type $j$ conditional on the observed types
in a sample of $n$ chromosomes from the population.   This
conditional probability under the coalescent does not seem to
be available explicitly (except under the unrealistic
\index{mutation!parent-independent mutation|)}parent-independent
mutation\index{mutation|)} assumption)---indeed an explicit
expression for this probability leads naturally to one for the
required likelihoods, and conversely.

Donnelly and Stephens exploited the structure of the discrete
representation of the Fleming--Viot
diffusion\index{Fleming, W. H.!Fleming--Viot diffusion} to approximate the
key conditional probabilities.  In effect, in the
\index{Donnelly, P. [Donnelly, P. J.]!Donnelly--Kurtz process}Donnelly--Kurtz
process they fixed the types on the first $n$ levels and ran the level $n+1$
process.  This led naturally to an approximation to the conditional
distribution of the $n+1$st sampled chromosome given the types of the first
$n$ chromosomes, which in turn leads naturally to importance sampling proposal
distributions.  As had been hoped, importance sampling under this family of
proposal distributions was considerably more efficient than under the
\index{Griffiths, R. C.!Griffiths--Tavar\'e approach (GT)|)}GT scheme
\cite{StD2000}.

There has been considerably more activity in the area of
inference under the coalescent over the last 10 years.  We will
not pursue this here, as our narrative will take a different
path. Interested readers are referred to \cite{Tav2004}\index{Tavar\'e, S.}
and \cite{Ste2007handbook}.

\section{The Li and Stephens model}\label{recombination}\index{LiN@Li, N.!Li and Stephens model|(}

As we have noted, statistical inference under the coalescent is hard.  From our
perspective, a key breakthrough came earlier this decade from
Li and Stephens \cite{LiS2003}.  Their idea was very simple, and it
turns out to have had massive impact.  Li and Stephens argued
that instead of trying to do inference\index{coalescence!inference|)}
under the coalescent one should appreciate that the coalescent is
itself only an approximation to reality, and that one might
instead do inference under a model which shares many of the
nice properties of the coalescent but also enjoys the
additional property that
\index{full likelihood inference@full-likelihood inference|)}full likelihood
inference is straightforward.

Li and Stephens changed the model.  Inference then became
a tractable problem.  What matters is how good these inferences
are for real data sets.  Although not obvious in advance, it
turns out that for a very wide range of questions, inference
under the Li and Stephens model works well in practice.

A forerunner to the Li and Stephens approach arose in
connection with the problem of estimating
\index{haplotype|(}haplotype phase from
\index{data!genotype data}genotype data.  Stephens, Smith, and Donnelly
\index{Stephens, M.}\index{Smith NJ@Smith, N. J.}\cite{SSD2001}
introduced an algorithm,
\index{phasing algorithm!phase@{\it PHASE}}{\tt PHASE}, in which the conditional
distribution underpinning the Donnelly--Stephens
\index{sampling!importance sampling|)}importance-sampling
\index{proposal distribution|)}proposal distribution was used directly in a
pseudo
\index{Gibbs, J. W.!Gibbs sampler}Gibbs sampler.
{\tt PHASE} has been widely used, and even
today provides one of the most accurate methods for
computational recovery of haplotype phase.  (Several recent
approaches aim to speed up computations to allow phasing of
genome-wide data sets, typically at a slight cost in accuracy e.g.\
\index{phasing algorithm!Beagle@{\it Beagle}}Beagle
\index{Browning, B. L.}\index{Browning, S. R.}\cite{browning2007rapid},
\cite{browning2009unified};
\index{phasing algorithm!FastPhase@\emph{FastPhase}}FastPhase
\index{Scheet, P.}\cite{scheet2006faf}; and
\index{phasing algorithm!impute@\emph{IMPUTE 2}}IMPUTE 2
\index{Howie, B. N.}\index{Marchini, J.}\cite{howie2009flexible}. See
\index{International HapMap Consortium}\cite{marchini2006comparison}
for a review of some of these methods.)

We now describe the Li and Stephens model.  For most modern
data sets it is natural to do so in the context of `SNPs'.
A \emph{SNP}, or
\index{single nucleotide polymorphism (SNP)|(}single nucleotide polymorphism,
is a position in the
\index{DNA (deoxyribonucleic acid)}DNA sequence which is known to vary
across chromosomes.
At the overwhelming majority of SNPs there will be exactly two
variants present in a population, and we assume this here.  For
ease, we will often code the variants as 0 and 1.  To simplify
the description of the model we assume haplotype data are
available.  This is equivalent to knowing the types at each SNP
separately along each of the two chromosomes in a \Index{diploid}
individual.   (Most experimental methods provide only the
unordered pair of types on the two chromosomes at each SNP,
without giving the additional information as to which variant
is on which chromosome.  As noted above, there are good
statistical methods for estimating the separate haplotypes from
these genotype data.)

It is convenient, and for many purposes most helpful, to
describe the Li and Stephens model via the conditional
probabilities it induces, and in particular by specifying the
probability distribution for the $n+1$st sampled chromosome
given the types of the first $n$ chromosomes. This in turn can
be described by a recipe for \index{simulation}simulating from this conditional
distribution. (We return below to a probabilistic aside on this
perspective.)

In effect, the Li and Stephens model simulates the $n+1$st
chromosome as an imperfect mosaic of the first $n$ chromosomes.
To simulate the $n+1$st chromosome, first pick one of the
existing $n$ chromosomes at random. At the first SNP
\index{copying|(}copy the type from the chosen chromosome, but with random
`error' in a way we will describe below.   With high
probability (specified below) the second SNP will be
probabilistically copied from the same chromosome.
Alternatively, the chromosome for copying at the second SNP
will be re-chosen, uniformly and independently. Having simulated
the type on the $n+1$st chromosome at the $k$th SNP, the
$k+1$st SNP will be copied from the same chromosome as the
$k$th SNP with high probability, and otherwise copied from a
chromosome chosen independently, and uniformly at random, from
the first $n$.  It remains to specify the probabilistic copying
mechanism: with high probability at a particular SNP, the value
of the $n+1$st chromosome will be the same as that on the
chromosome being copied, otherwise it will have the opposite
type.

The connection with the coalescent comes from the following.  Consider the
position of the first SNP on the $n+1$st chromosome.  Ignoring coalescences
amongst the first $n$ sampled chromosomes at this position, the ancestry of
the $n+1$st sampled chromosome coalesces with exactly one of the lineages
leading to the $n$ sampled chromosomes.  Ignoring mutation\index{mutation|(},
the type of the $n+1$st chromosome at this position will be the same as the
type on the chromosome with which its ancestry coalesces. To incorporate
mutation one allows mis-copying of the ancestral type\index{type|)}.  This
mis-copying is an oversimplification of the effect of mutation on
the \Index{coalescent tree} at this position.  Now, moving along the $n+1$st
chromosome, there will be a segment, up to the
first \index{recombination|(}recombination event in the relevant history,
which shares the same ancestry (and so is copied from the same one of the
sampled chromosomes).  The effect of recombination is to follow different
ancestral chromosomes and this is mimicked in the Li and Stephens approach by
choosing a different chromosome from which to copy.  The probabilities of this
change will depend on the recombination rates between the SNPs, and in a
coalescent also on $n$, because coalescence of the \Index{lineage} of the
$n+1$st chromosome to one of the other lineages happens faster for larger $n$.

We now describe the model more formally.  Suppose that $n$ chromosomes (each
of which can be thought of as a haplotype) have been \index{sampling|(}sampled
from the population, where the $j$th haplotype has the SNP information at $l$
SNPs, $c^j = \{c_1^j, c_2^j, \dots, c_l^j\}$. Let us call this set of
chromosomes $C$.  Now suppose an additional chromosome $i$ has been sampled
and has SNP information $h^i =
\{h_1^i, h_2^i, \dots, h_l^i\}$.  We seek to determine the
probability of sampling this chromosome, based on its SNP
haplotype and the SNP haplotypes of the previously sampled
chromosomes $C$.  The model takes as input fine-scale estimates
of recombination rates in the region: $r = \{r_0, r_1,\dots,
r_l\}$ where $r_{j + 1} - r_j$ is the average rate of \Index{crossover}
per unit physical distance per \Index{meiosis} between sites $j$ and $j
+ 1$ times the physical distance between them.  We set $r_0 =
0$.  We obtain this
\index{genetic map}map from elsewhere (for example
\index{International HapMap Consortium}\cite{IHC2007})
rather than estimating it for ourselves\footnote{In fact, Li
and Stephens developed their model precisely for estimating the
genetic map, but for our purposes we wish to utilize the model
for inference in other settings and thus we utilize a known genetic map.}.  Note
that the SNPs (and the map) are ordered by the position of the
SNP (or map point) on the chromosome (for convenience we refer
to the first SNP position as the leftmost position and the
$l$th SNP position as the rightmost). We define the per-locus
recombination probability $\rho_s = 1 - \exp(-4N_e(r_{s+1} -
r_s)/n)$ and then define transition probabilities for a
\index{Markov, A. A.!Markov chain}Markov chain on $\{1, 2, \ldots, n\}$
from state
$j$ (indicating that it is the $j$th haplotype of those that
have been previously sampled that is `parental') at position
$s$ to state $k$ at position $s + 1$: \begin{equation}
\label{eq:transition} q(j_s, k_{s + 1}) =
\begin{cases} 1 - \rho_s + \rho_s/n, & j = k, \\
\rho_s/n, & j \neq k, \\
    		\end{cases}
\end{equation} where $N_e$ is the so-called
\Index{effective population size}, a familiar quantity in population genetics
models.  Equation (\ref{eq:transition}) is related to the fact that recombination
events occur along the sequence as a Poisson process.  Here we use the Poisson
rate $4N_e(r_{s+1} - r_s)/n$.  Given the rate, the probability that there is
no recombination between sites $s$ and $s + 1$ is $\exp(-4N_e(r_{s+1} -
r_s)/n) = 1 - \rho_s$.  The probability of at least one recombination between
sites $s$ and $s + 1$ is thus $\rho_s$.  In this case the model has the
assumption that it is equally likely that the recombination occurs with any of
the $n$ sampled haplotypes.  In particular, as $\rho_s$ incorporates the
probability that multiple recombinations occur between sites $s$ and $s + 1$,
the first case in Equation (\ref{eq:transition}) includes a $\rho_s$ term to
allow for the possibility that the same haplotype is parental at each site $s$
and $s + 1$ even when one or more recombinations have occurred.

We define the copying probabilities in terms of the
\index{mutation!rate}`population mutation rate' for the
given sample size
($\theta$, defined below), another familiar quantity from
population genetics.  The mismatch (or not) between the SNP
allele of the $j$th `parent' chromosome at SNP $s$, $c_s^j$,
and the SNP \Index{allele} of the $i$th additional `daughter'
chromosome, $h_s^i$, is defined as \begin{equation}
\label{eq:emission} e(h_s^i, c_s^j) = 		\begin{cases}
\frac{n}{n + \theta} + \frac{1}{2} 			\frac{\theta}{n +
\theta}, & h_s^i = 			c_s^j , \\ \frac{1}{2}
\frac{\theta}{n + 			\theta}, & h_s^i \neq c_s^j. \\
\end{cases} \end{equation} Notice that as $\theta \rightarrow
\infty$ the alleles $0$ and $1$ at any given site become
equally likely.  Equation (\ref{eq:emission}) is motivated by a
similar coalescent argument to that used for the transition
probabilities above.  In this case the probability of no
mutations occurring at the site $s$ is $n/ (n + \theta)$ and
thus the probability of at least one mutation occurring at $s$
is $\theta / (n + \theta)$.   It is possible to allow
for the population mutation rate to vary sitewise if it is
necessary to account for known variable mutation rates.

The particular form of the transition and copying probabilities
in the model follow from the informal coalescent arguments
given four paragraphs above (\cite{LiS2003}).  We
noted that the recombination probabilities typically come from
available estimates.  It turns out that the accuracy of these estimates can be important in applications of the model.  In contrast, such applications are generally not especially sensitive to the exact value of $\theta$ used.  Thus, for the mutation probabilities, we follow Li and Stephens and set
\begin{equation}
\label{eq:theta} \theta = \left(\sum_{z = 1}^{n - 1}
\frac{1}{z}
	\right)^{-1}.
\end{equation}

\begin{figure}
% \centering
  \includegraphics[width = \textwidth]{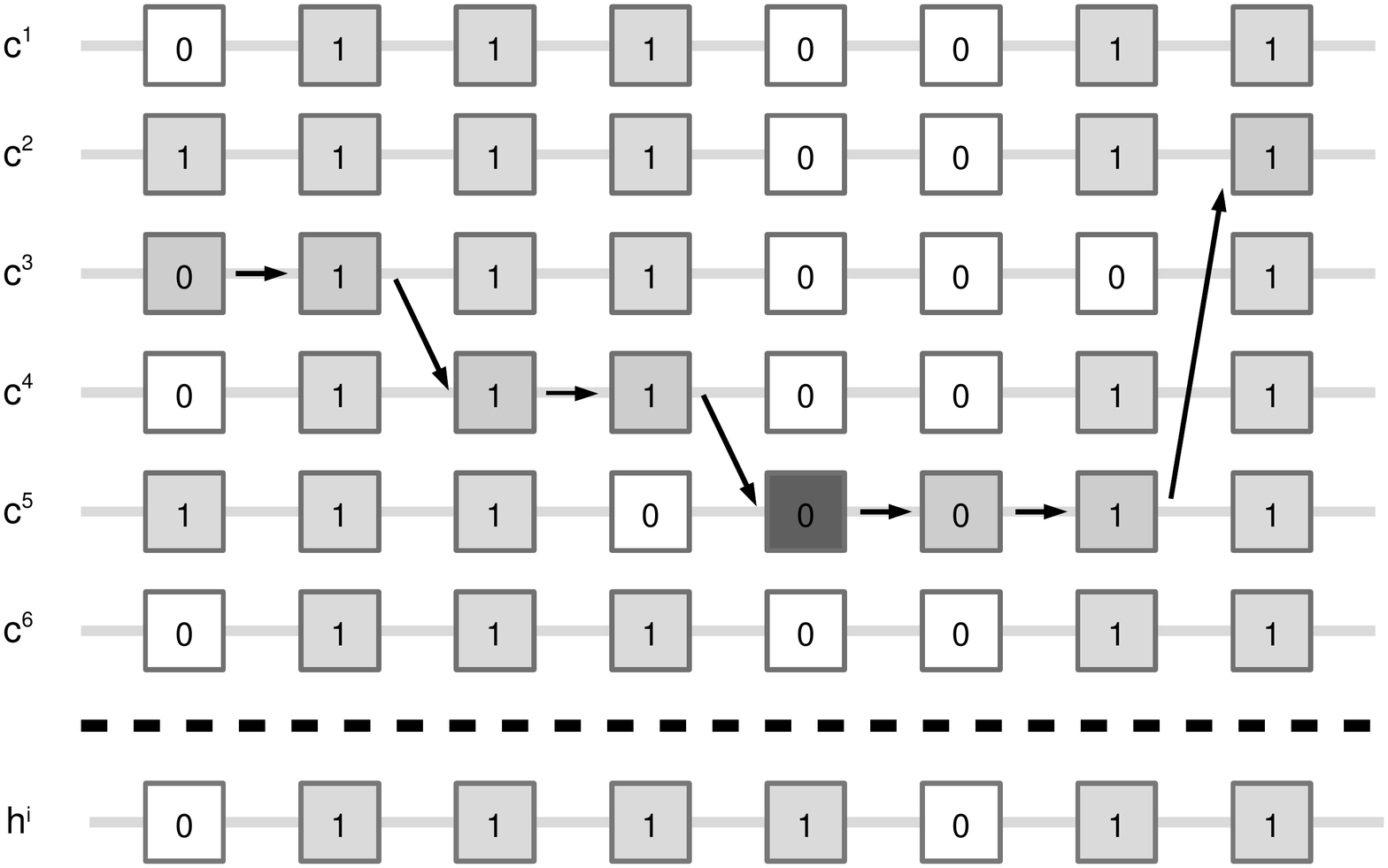}
  \caption{A pictorial representation of the Li and Stephens model for the
  simulation of a new haplotype.  Here we have sampled sequences $c^1$, \dots,
  $c^6$ and seek to simulate a new haplotype $h^i$.  The model first simulates
  a `path' through the sampled sequences, indicated in the figure by the
  arrows, which show which of the $c^i$ is copied at each SNP locus.
  Recombination with another parental sequence is indicated by changing the
  sequence being copied (for example, between the second and third loci).  The
  dark box on the path at the fifth locus indicates that a mutation\index{mutation|)} occurred
  in the copying of that locus.}
%  \caption{A pictorial representation of the Li and Stephens calculation for a
%  single path.  Here we have sampled sequences $c^1, \dots, c^6$ and seek to
%  calculate the conditional probability of sampling $h^i$ by summing the
%  probabilities obtained over all possible paths.  The arrows indicate the
%  `path' taken through the sampled sequences, indicating which of the $c^i$ has
%  the `parental' type at a given locus.  Recombination with another parental
%  sequence is indicated by changing the sequence being copied (for  example,
%  between the second and third loci).  Green boxes indicate that the copying is
%  faithful (i.e.\ there is no mutation), whereas red boxes indicate a mutation
%  has occurred.}
 \label{condprob}
\end{figure}

We can view the Li and Stephens process as defining a path through the
previously sampled sequences $C$. This is illustrated in
Figure \ref{condprob}.

A key feature of the conditional probabilities just described
for the Li and Stephens model is that they take the form of a
\index{Markov, A. A.!hidden Markov model (HMM)}hidden Markov model (HMM)
\index{Durbin, R.|(}\index{Eddy, S. R.|(}\index{Krogh, A.|(}\index{Mitchison, G.|(}\cite{durbin1998biological}.  The path through
the sampled chromosomes is a Markov chain on the set $\{1, 2,
\ldots, n\}$ which indexes these chromosomes: the value of the
chain at a particular SNP specifies which chromosome is being
copied at that SNP to produce the new chromosome.  In the
language of HMMs, the \emph{transition probabilities} specify the probability
of either continuing to copy the same chromosome or choosing another
chromosome to copy, and the
\index{emission probability}\emph{emission probabilities} specify
the probability of observing a given value at the SNP on the
new chromosome given its type\index{type|(} on the chromosome being copied.
The latter have the simple form that with high probability the new
chromosome will just \index{copying|)}copy the type from the chromosome being
copied, with the remaining probability being for a switch
to the opposite type from that on the chromosome being copied.  Viewed this way, the transition probabilities relate to the recombination process and the emission probabilities to the mutation process.
The reason that the HMM structure is crucial is that very
efficient \index{algorithm}algorithms are available for calculations in this
context.  For example, under the Li and Stephens model, given values for all the SNPs on the
$n+1$st chromosome, and good computational resources, one can
calculate the conditional probability of each possible path
through the first $n$ chromosomes, or of the \Index{maximum likelihood}
path (see
\index{Durbin, R.|)}\index{Eddy, S. R.|)}\index{Krogh, A.|)}\index{Mitchison, G.|)}\cite{durbin1998biological} for details).

Not only is the Li and Stephens model tractable, in a way that
the coalescent is not, but it turns out that its use for
inference in real populations has been very successful.
Examples include inference of \index{haplotype|)}haplotype phase
\index{Scheet, P.}\index{Stephens, M.}\cite{scheet2006faf},
\cite{stephens2005adl},
\index{LiY@Li, Y.}\index{Ding, J.}\index{Abecasis, G. R.}\cite{mach1},
\index{Willer, C. J.}\cite{mach2}; inference of
fine-scale \index{recombination|)}recombination rates \cite{LiS2003};
imputation of unobserved \index{data!genotype data}genotype data
\index{Marchini, J.}\index{Howie, B. N.}\index{Myers, S. R.}\index{McVean, G. A. T.}\cite{marchini2007nmm},
\index{Servin, B.}\cite{servin2007iba}; and imputation
of classical
\index{human leukocyte antigen (HLA)}HLA types\index{type|)} from SNP data
\index{McVean, G. A. T.}\cite{Leslie:2008ly}.
It seems that the model captures enough of the features of real
data to provide a good framework for inference.  Because
\index{coalescence!inference}inference under the coalescent is impossible,
it is not known
whether this would have properties which are better or worse
than those under Li and Stephens, though we would not expect
major differences.

We have specified the Li and Stephens model via its conditional
probabilities, and these are what is crucial in most
applications.  But there is a probabilistic curiosity which is
worth mentioning.  One could calculate the probability, under
the model, for a particular configuration of $n$ chromosomes
via these conditional probabilities: for a particular ordering,
this would simply be the product of the marginal probability
for the first chromosome, the Li and Stephens conditional
probability for the second chromosome given the first, the Li
and Stephens conditional probability for the third given the
first and second, and so on.  In fact, Li and Stephens
themselves referred to their model as the product of
approximate conditionals, or
\index{product of approximate conditionals (PAC)}PAC, \Index{likelihood}.  The
probabilistic curiosity is that in this formulation the
likelihood, or equivalently sampling probabilities, in general
depend on the order in which the chromosomes are considered.
One could solve the problem by averaging over all possible
orders, or approximately solve it by averaging over many
orders, the approach adopted by Li and Stephens.  But for most
applications, including the one we describe below, it is the
conditional distributions which matter, either in their own
right (e.g.
\index{McVean, G. A. T.}\cite{Leslie:2008ly}) or for use in what resembles a
\index{Gibbs, J. W.!Gibbs sampler}Gibbs sampler.
This latter approach gives rise to another curiosity: one can
write down, and implement, an \Index{algorithm} using the Li and
Stephens conditionals as if it were a Gibbs sampler, even
though the conditionals do not obviously correspond to a proper
joint distribution.  These algorithms (of which
\index{phasing algorithm!phase@{\it PHASE}}{\tt PHASE} was
perhaps the first) often perform extremely well in practice,
notwithstanding the gap in their probabilistic pedigree, an
observation which might warrant further theoretical
investigation.

\section{Application: modelling population structure}\label{conclusion}

\index{population structure|(}In this section we describe an extension of
the Li and Stephens model appropriate for
\index{population structure!geographical population structure}geographically
structured populations, and then show how
\index{statistical inference}inference under the model performs well on real
data.

Real populations often consist of genetically distinct
subpopulations, and there has been considerable interest in the
development of statistical methods which detect such
subpopulations, and assign sampled individuals to them, on the
basis of population genetics data
\index{Bowcock, A. M.}\index{Ruiz-Linares, A.}\index{Tomfohrde, J.}\cite{Bowcock:1994wm},
\index{Price, A. L.}\cite{Price:2006rc},
\index{Patterson, N.}\index{Reich, D.}\cite{Patterson:2006xy},
\index{Pritchard, J. K.}\index{Stephens, M.}\cite{PSD2000},
\index{Falush, D.}\cite{FSP2003},
\index{Corander, J.|(}\index{Waldmann, P.|(}\cite{corander2004bep},
\index{Dawson, K. J.|(}\index{Belkhir, K.|(}\cite{dawson2001bai}.  Understanding
population structure is of interest in \Index{conservation biology},
\Index{human anthropology}, \Index{human disease genetics}, and
\Index{forensic science}.  It is important to detect hidden
population structure for \Index{disease mapping} or studies of
\index{gene!flow}gene flow, where failure to detect such structure may result in
misleading inferences.  Population structure is common amongst
organisms, and is usually caused by subpopulations forming due
to geographical subdivision.  It results in genetic
differentiation---frequencies of variants, (called
\index{allele}\emph{alleles} in genetics)
that differ between subpopulations.  This may be due to
\Index{natural selection} in different environments,
\Index{genetic drift} (stochastic
fluctuations in population composition) in distinct
subpopulations, or chance differences in the genetic make up of
the founders of subpopulations \cite{HaC1997}.

Model-based approaches to detecting and understanding
population structure have been very successful
\index{Pritchard, J. K.}\index{Stephens, M.}\cite{PSD2000},
\index{Falush, D.}\cite{FSP2003},
\index{Corander, J.|)}\index{Waldmann, P.|)}\cite{corander2004bep},
\index{Dawson, K. J.|)}\index{Belkhir, K.|)}\cite{dawson2001bai}.  Broadly, one
specifies a statistical model describing data from different
subpopulations and then performs inference under the model.
These models are examples of the statistical class of
\index{mixture model|(}\emph{mixture models}, in which
observations are modelled as coming
from a (typically unknown) number of distinct classes.   In our
context the classes consist of the subpopulations, and what is
required is a model for the composition of each, and the way in
which these are related.   Model-based approaches to
understanding population structure have several natural
advantages.  Results are readily interpretable, and, at least for
\index{Bayes, T.!Bayesian inference|(}Bayesian inference procedures, they
provide a coherent assessment of the uncertainty associated with the
assignment of individuals to subpopulations, and the assessment of the number
of populations in the sample.

Where the data consist of SNPs taken from distinct regions of
the \Index{genome} it is natural to model these as being independent of
each other, within subpopulations, and it remains only to model
the frequency of the alleles at each SNP in each subpopulation,
and the joint distribution of these across subpopulations.
This approach was taken by Pritchard, Stephens and Donnelly
\cite{PSD2000} and implemented in the program 
\index{structure program@\emph{STRUCTURE} program}{\tt STRUCTURE} which has
been successfully used in a variety of applications.

In some contexts the population in question arose from the
mixing, or admixing as it is known in genetics, of distinct
populations at some time in the relatively recent past.
African-American populations are an example of \Index{admixture},
in this case between Caucasian and African populations.  Such
admixture is well known to lead to correlations between SNPs
over moderately large scales (say 5--10 million bases) across
chromosomes, known as admixture
\index{linkage disequilibrium|(}linkage disequilibrium.
Falush, Stephens and Pritchard \cite{FSP2003} adapted the model underlying {\tt
STRUCTURE} to incorporate these correlations, with the new
model also having been widely applied
\index{Falush, D.}\index{Pritchard, J. K.}\index{Stephens, M.}\cite{FSP2003},
\index{Wirth, T.}\cite{falush2003traces}.

SNPs which are close to each other on the chromosome\index{chromosome|)} exhibit
correlations, known in genetics as
\index{linkage disequilibrium|)}\emph{linkage disequilibrium}, due to their
shared ancestral history. Both the coalescent\index{coalescence|)} with
\index{recombination}recombination and the Li and Stephens model explicitly
capture these correlations for a single randomly-mating population.  To employ
a model-based approach to population structure for nearby SNPs with linkage
disequilibrium, one needs to model these correlations within and between
subpopulations. We introduce such a model below as a natural extension of the
Li and Stephens model.

For simplicity throughout, we describe our model assuming the
\index{haplotype|(}haplotypes of the sampled individuals are known, so we assume
that the phase of the data is known, either directly or by
being inferred using a \Index{phasing} method such as
\index{Stephens, M.}\index{Smith NJ@Smith, N. J.}\cite{SSD2001} or
\index{Scheet, P.}\cite{scheet2006faf}.  Given the accuracy of statistical
phasing methods, this is a reasonable assumption in most
cases, but as we note below, the assumption can be dropped, at
some computational cost.

Suppose we have
\index{DNA (deoxyribonucleic acid)}DNA sequence data (SNPs) for $N$
\Index{haploid} individuals (or $N$ phased haplotypes) sampled from $K$
populations at $L$ loci.  Call these data $H = \{h_1, \dots,
h_N\}$, where the index $i$ represents individual $i$.  Define
the population assignment of individual $i$ to be the discrete
random variable $Z_i$ which can take values in $\{1, \dots,
K\}$.  In contrast to the previous methods, however, we make no
assumption that the SNPs are independent (or merely loosely
dependent as is the case for \cite{FSP2003}) i.e.\ we
explicitly deal with linkage disequilibrium.  In order to model
linkage disequilibrium, by applying the model of Li and
Stephens \cite{LiS2003}, we require a \Index{genetic map} of the region
covering the $L$ loci.  We assume we have such a map, obtained
from other sources (e.g.\ the program
\index{ldhat@\emph{LDhat} program}{\tt LDhat}
\index{McVean, G. A. T.}\index{Awadalla, P.}\index{Fearnhead, P.}\cite{MAF2002},
\index{Myers, S. R.}\cite{MMHDBR2004},
\index{Auton, A.}\cite{auton2007recombination} or from applying the Li and
Stephens method for estimating recombination rates \cite{LiS2003}).  In
specifying the model we assume we know the value of $K$.

We wish to allocate sequences to populations based on the data
$H$. As is natural in
\index{mixture model|)}mixture models, we do so in a
\index{Bayes, T.!Bayesian inference|)}Bayesian framework, via
\index{Gibbs, J. W.!Gibbs sampler}Gibbs sampling (or to be more precise, as we
note below, via pseudo-Gibbs sampling) over the unobserved
allocation variables, the $Z_i$.   To do so we need to specify
the conditional distribution of $Z_i$ given the data $H$ and
the values of the allocation variables $Z$ for individuals
other than individual $i$.

We specify the prior distribution on the $Z_i$ to be uniform, i.e.
\[
\prob(Z_i = 1) = \prob(Z_i = 2) = \cdots = \prob(Z_i = K) = \frac{1}{K}
\]
for every $i = 1$, \dots, $N$.  It is a simple matter to include an informative
prior if required.  Furthermore, we assume that in the absence of any data
$h_i$, $Z_i$ is independent of all of the $Z_j$ and $h_j$ for $i \neq j$.

We first informally describe the model in the special case in
which the ancestry of any chromosome\index{chromosome|(} only ever involves
chromosomes in its own population.  In effect this means that
there has been no \Index{migration} between subpopulations over the
time back to the common ancestors of sampled chromosomes.  In
this special case, we can think of calculating the conditional
distribution as follows. (We describe the calculation of the
conditional distribution of $Z_1$ given $H$ and the other
$Z_i$.)  Suppose we know the current population assignments in
our sampler of all of the haplotypes.  We wish to update the
assignment of haplotype 1 based on the assignments of all of
the other haplotypes.  First, we remove haplotype 1 from the
population it is currently assigned to.  Then, we calculate the
Li and Stephens conditional probability that $h_1$ would be the
next sampled haplotype from each of the $K$ populations.  We
normalize these probabilities by dividing by their sum and draw
the new population assignment of haplotype 1 from the
multinomial distribution with these normalized probabilities as
the population weights.

We now introduce the general version of our new model, which
extends the simple Li and Stephens models above to incorporate
migration or recent \Index{admixture}, and use this as the basis for
our method for classifying individuals into populations.
Importantly, the model reduces to the Li and Stephens model in
certain limiting cases.  The method uses SNP haplotype data and
explicitly models correlations between SNP loci using the model
and estimates of
\index{recombination|(}recombination rates between the loci.  We test
our method on a sample of individuals from three continents and
show that it performs well when classifying individuals to
these continental populations.

We extend the Li and Stephens \cite{LiS2003} model to
incorporate admixture or inter-population migration by adding
another parameter to the model, represented by $\alpha$, which
models the extent of shared ancestry between populations. This
parameter attempts to capture the contribution to an
individual's ancestry which is derived from a population other
than its own.  It does not have a natural interpretation in
terms of the dynamics of the populations forward in time.  In
this sense it has more in common with parameters in statistical
models than in probability models.  This extended model is
perhaps best understood by considering the \index{copying|(}`copying path'
representation of the Li and Stephens model (see
Figure \ref{condprob}).  For simplicity we present the extended
model for the case of two populations (i.e.\ $K = 2$) with a
fixed value of $\alpha$.  We discuss further generalizations
after the initial presentation of the model.

As in the special case above, the model allows calculation of
the probability that a particular haplotype is sampled from a
given population.  Normalising these probabilities then gives
the probabilities for resampling the allocation variables.

The extension of Li and Stephens relates to
recombination, or in the sense of Figure \ref{condprob}, to the step when
there is a potential change to the chromosome being copied.
When such a potential change occurs in the simple model a new
chromosome is chosen to be copied uniformly at random
(including the chromosome currently being copied).   In the
 extension described above, this remains the case, but
with the new chromosome allowed to be chosen only from the
population being considered.  Our generalisation allows the
possibility that the copying process could switch to a
chromosome in a different population, and the parameter
$\alpha$ controls the relative probability of this.

We now give the details of the model we use.   Suppose that
$n_1$ and $n_2$ (where $n_1 + n_2 = n$) chromosomes have been
sampled from populations $1$ and $2$ respectively, where the
$j$th haplotype in the total population has the SNP information
at $l$ SNPs, $c^j = \{c_1^j, c_2^j, \dots, c_l^j\}$.   Let us
call this set of chromosomes $C = C_1 \cup C_2$, where $C_1$
are the chromosomes sampled from Population 1, and $C_2$ from
Population 2.  Now suppose an additional chromosome $i$ has
been sampled and has SNP information $h^i = \{h_1^i, h_2^i,
\dots, h_l^i\}$.  Without loss of generality we seek to
determine the probability of sampling this chromosome from
Population $1$, based on its SNP haplotype and the SNP
haplotypes of the previously sampled chromosomes $C$.   We
define the per locus recombination probability $\rho_s = 1 -
\exp(-4N_e(r_{s+1} - r_s)/(n_1 + \alpha_1 n_2))$ and then
define the transition probabilities from state $j$ (indicating
that it is the $j$th haplotype of those that have been
previously sampled that is `parental') at position $s$ to state
$k$ at position $s + 1$: \begin{equation}
\label{eq:alpha_transition} q(j_s, k_{s + 1}) = \begin{cases} 1
-   \rho_s + \frac{\rho_s}{n_1 + \alpha_1     n_2}, & j = k,\ j
\in C_1,\\
  \frac{\rho_s}{n_1     + \alpha_1 n_2}, & j \neq k,\ j \in C_1,\   k \in C_1,
  \\
  \alpha_1   \frac{\rho_s}{n_1 + \alpha_1 n_2}, & j   \neq k,\ j \in C_1,\ k \in
  C_2, \\
  1 - \rho_s +   \alpha_1 \frac{\rho_s}{n_1 + \alpha_1     n_2}, & j = k,\ j \in
  C_2, \\
  \frac{\rho_s}{n_1     + \alpha_1 n_2}, & j \neq k,\ j \in C_2,\   k \in C_1,
  \\
  \alpha_1   \frac{\rho_s}{n_1 + \alpha_1 n_2}, & j   \neq k,\ j \in C_2,\ k \in
  C_2, \\
    		\end{cases}
\end{equation}
where $N_e$ is the
\Index{effective population size}.  Notice that when $\alpha_1 = 0$
the transition probabilities so defined reduce to the Li and Stephens
transition probabilities for a single sample from Population 1.  Also note that
when $\alpha_1 = 1$ we have effectively combined the two samples into a single
sample from one population and again we obtain the Li and Stephens transition
probabilities for this case.

We define the
\index{emission probability}emission probabilities in terms of the
`population mutation rate'\index{mutation!rate}
for our given sample size, where in this case our `sample size' is $n_1 +
\alpha_1 n_2$,
\begin{equation}
  \label{eq:alpha_theta} \theta =   \left(\sum_{z = 1}^{n_1 - 1}
  \frac{1}{z} + \sum_{z = n_1}^{n_1 +   n_2 - 1} \frac{\alpha_1}{z}
  \right)^{-1}.
\end{equation}
Again, we obtain the desired Li and Stephens population mutation rates for the
$\alpha_1 = 0$ and $\alpha_1 = 1$ cases.  We then define the mismatch (or not)
between the SNP allele of the $j$th `parent' chromosome at SNP $s$, $c_s^j$,
and the SNP allele of the $i$th additional `daughter' chromosome, $h_s^i$:
\begin{equation}
\label{eq:alpha_emission} e(h_s^i, c_s^j) =
		\begin{cases} \frac{n_1 + \alpha_1 n_2}{n_1 + \alpha_1 n_2 + \theta} +
\frac{1}{2}
			\frac{\theta}{n_1 + \alpha_1 n_2 + \theta}, & h_s^i =
			c_s^j, \\ \frac{1}{2} \frac{\theta}{n_1 + \alpha_1 n_2 +
			\theta}, & h_s^i \neq c_s^j. \\
		\end{cases} \end{equation} As before, notice that these
emission probabilities reduce to the analogous
\index{LiN@Li, N.!Li and Stephens model|)}Li and Stephens
emission probabilities for the cases $\alpha_1 = 0$ and
$\alpha_1 = 1$.

\begin{figure}
% \centering
  \includegraphics[width = \textwidth]{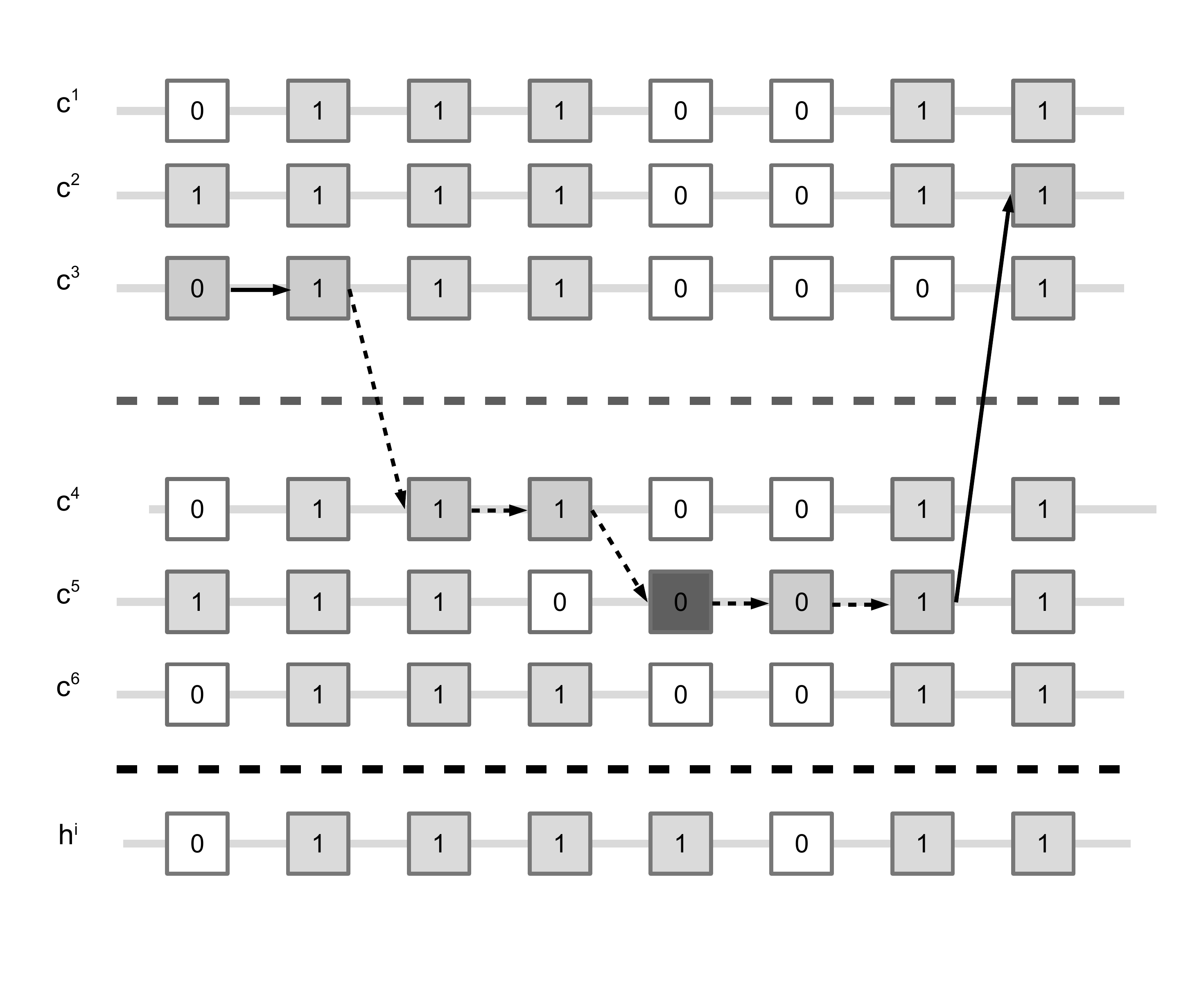}
  \caption{A pictorial representation of the calculation for a single path
  using the new model.  Here we have sampled sequences $c^1$, \dots, $c^6$ from
  two populations, represented by the first three rows and the second three
  rows in the figure, respectively.  We seek to calculate the probability of
  sampling $h^i$ (seventh row) from the first population by summing the
  probabilities obtained over all possible paths.  We illustrate this with a
  single path.  The arrows indicate the `path' taken through the sampled
  sequences, indicating which of the $c^i$ has the `parental' type\index{type} at a given
  SNP locus.  Recombination with another parental sequence is indicated by
  changing the sequence being copied (for example, between the second and
  third loci).  The dark box on the path at the fifth locus indicates that a
  \Index{mutation} occurred in the copying of that locus.  A dashed arrow indicates
  that the \index{copying|)}copying at the next position is taking place with a sequence not in
  the population we are interested in (in this case the first population) and
  thus the \emph{recombination term} in the model is scaled by the $\alpha$
  parameter for these terms.}
%  \caption{A pictorial representation of the calculation for a single path
%    using the new model.  Here we have sampled sequences $c^1, \dots, c^6$
%    from two populations (represented by the green and blue borders on the
%    squares).  We seek to calculate the probability of sampling $h^i$
%    (represented by the orange borders on the squares) from the first (green)
%    population by summing the probabilities obtained over all possible paths.
%    We illustrate this with a single path.  The arrows indicate the `path'
%    taken through the sampled sequences, indicating which of the $c^i$ has the
%    `parental' type at a given locus.  Recombination with another parental
%    sequence is indicated by changing the sequence being copied (for example,
%    between the second and third loci).  Green filled boxes indicate that the
%    copying is faithful (i.e.\ there is no mutation), whereas red boxes
%    indicate a mutation has occurred.  A dashed arrow indicates that the
%    copying at the next position is taking place with a sequence not in the
%    population we are interested in (i.e.\ in this case the blue population)
%    and thus the \emph{recombination term} in the model is scaled by the
%    $\alpha$ parameter for these terms.}
 \label{sec:LD_condprob}
\end{figure}

As was the case for a single population, we can view this
process as defining a path through the previously sampled
sequences $C$. This is illustrated in
Figure \ref{sec:LD_condprob}. 

Using this model we proceed as before.  To calculate the conditional
probability of observing the additional haplotype we sum over all possible
paths through the potential parental chromosomes $C$.  We use the forward
\index{algorithm|(}algorithm which gives a computationally efficient means of
performing the
required summation
\index{Durbin, R.}\index{Eddy, S. R.}\index{Krogh, A.}\index{Mitchison, G.}\cite{durbin1998biological}. For each of the $n$
previously sam\-pled chromosomes, we
initialise the forward algorithm:
\begin{equation}
\label{eq:alpha_fwdinit}
	f_1^j = \begin{cases}
			\frac{1}{n_1 + \alpha_1 n_2} \times e(h_1^i, c_1^j), & j \in C_1,
\\
			\frac{\alpha_1}{n_1 + \alpha_1 n_2} \times e(h_1^i, c_1^j), & j \in
C_2.
		\end{cases}	
\end{equation}
The forward
\index{algorithm|)}algorithm moves along the sequence such that at each SNP $s$, where
$1 < s \leq l$,
\begin{equation}
\label{eq:alpha_fwd} f_s^j = e(h_s^i, c_s^j) \sum_{k = 1}^{n} f_{s - 1}^k
	\times q(k_{s - 1}, j_s).
\end{equation}

The probability of observing the SNP configuration of the additional chromosome
is given by
\begin{equation}
  \label{eq:alpha_likelihood} \hat{\pi}(h^i | C, \theta, \rho) = \sum_{j =
  1}^{n} f_l^j.
\end{equation}

A simple extension for $K > 2$ treats the current population of
interest as one population and all other populations are
combined to form the second population, thus reducing this case
to the $K = 2$ case.  This is what we consider here.  To
further generalize the model one may define a matrix of
$\alpha$ parameters for each ordered pair of populations.
Inference for the number of populations represented in the
sample ($K$) may be performed using the method of
\index{Pritchard, J. K.}\index{Stephens, M.}\cite{PSD2000} or by other
methods.  We focus here only on the
problem of classifying individuals where the number of
subpopulations is known.

As noted above, it is a slight abuse of terminology to refer to
the process just described as
\index{Gibbs, J. W.!Gibbs sampler}Gibbs sampling.   In this case,
we cannot be certain that our approximations do in fact
correspond to a proper joint probability.  Nonetheless, there
are successful precedents for this use of `pseudo-Gibbs'
sampling, notably the program
\index{phasing algorithm!phase@{\it PHASE}}{\tt PHASE}
\index{Stephens, M.}\index{Smith NJ@Smith, N. J.}\cite{SSD2001},
\cite{StD2003} which has proved successful 
for inferring haplotypes from genotypes. Furthermore, our
method is successful in practice.  Given these caveats, we
continue our abuse of terminology throughout.

It is also possible to set the proportion of the sample coming from each
population as a parameter and update this value at each step.  This would
allow us to deal easily with populations of different sizes in a
straightforward manner, although we have not implemented this extension.
Extending the model to incorporate genotypes rather than haplotypes is also
straightforward in principle, for example by analogy
with
\index{Marchini, J.}\index{Howie, B. N.}\index{Myers, S. R.}\index{McVean, G. A. T.}\cite{marchini2007nmm}.

Our model applies when SNPs are sampled from the same small
chromosomal region. In practice, it is more likely that data of
this type will be available for a number of different,
separated, chromosomal regions.  In this context we apply the
model separately to each region and combine probabilities
across regions multiplicatively.    We implement this by
setting $\rho_s = 1$ in Equation (\ref{eq:alpha_transition}) for
the transition from the last SNP in one region to the first SNP
in the next region (the order of the regions is immaterial).

We tested the method on phased
\index{data!HapMap|(}data available from the Phase II HapMap
\index{International HapMap Consortium}\cite{IHC2007}.  In particular the
data consisted of SNP
haplotypes from samples from four populations: Yoruba ancestry
from Ibadan in Nigeria (YRI); European ancestry from Utah
(CEU); Han Chinese ancestry from Beijing (CHB); and Japanese
ancestry from Tokyo (JPT).  We use the SNP haplotype data
available from the HapMap and the
\index{recombination|)}recombination map estimated
in that study from the combined populations.  The YRI and CEU
HapMap samples are taken from 30 parent offspring trios in each
case.  Of these we used the data from the parents only, giving
60 individuals (120 haplotypes) in each population.  The
Chinese and Japanese samples derive from 45 unrelated
individuals (90 haplotypes) in each case.  Following common
practice, we combine the Chinese and Japanese samples into a
single `analysis panel', which we denote by CHB+JPT.  Thus our
sample consists of a total of 420 haplotypes deriving from
three groups.

From these data we selected 50 independent regions each of 50 kilobases in
length (i.e.\ regions not in
\Index{linkage disequilibrium} with each other).  Within
these regions we selected only those SNPs that are variable in
all three populations and have a minor allele frequency of at least $0.05$.
A summary of the data is given in Table \ref{tab:DataSummary_MAF_0.05}.

%MAF = 0.05
\begin{table}
%\centering
\begin{minipage}{100mm}
\caption{Summary of the data: The data consist of HapMap samples for 420
  haplotypes (120 CEU, 180 CHB+JPT, 120 YRI) sampled from 50 independent
  regions of $\sim$50kb each, taken across all autosomes.  SNPs with a minor
  allele frequency of less than 0.05 in any of the population groups have been
  excluded.  For each region the number of SNPs segregating in all populations
  is shown, as well as the region size, which is the distance between the
  first and last SNP in the region.}
\label{tab:DataSummary_MAF_0.05}
\end{minipage}
\begin{tabular}{rrrr|}
	\begin{sideways} Chromosome Number \end{sideways} & \begin{sideways} Region ID \end{sideways} & \begin{sideways} Number of SNPs \end{sideways} & \begin{sideways} Region Size (bp) \end{sideways} \\
	\hline
         1 &          0 &          6 &      38600 \\
         2 &          1 &         42 &      48726 \\
         3 &          2 &         20 &      46100 \\
         4 &          3 &         49 &      49031 \\
         5 &          4 &         82 &      43511 \\
         6 &          5 &         53 &      49377 \\
         7 &          6 &         37 &      48387 \\
         8 &          7 &         86 &      49212 \\
         9 &          8 &         55 &      49700 \\
        10 &          9 &         59 &      49620 \\
        11 &         10 &         29 &      46888 \\
        12 &         11 &         76 &      49762 \\
        13 &         12 &         30 &      48056 \\
        14 &         13 &         23 &      49648 \\
        15 &         14 &         33 &      47289 \\
        16 &         15 &         85 &      47207 \\
        17 &         16 &         38 &      47062 \\
        18 &         17 &         61 &      48001 \\
        19 &         18 &         33 &      48359 \\
        20 &         19 &         44 &      48119 \\
        21 &         20 &         45 &      49195 \\
        22 &         21 &          1 &          0 \\
\end{tabular}
\begin{tabular}{rrrr|}
	\begin{sideways} Chromosome Number \end{sideways} & \begin{sideways} Region ID \end{sideways} & \begin{sideways} Number of SNPs \end{sideways} & \begin{sideways} Region Size (bp) \end{sideways} \\
	\hline
         1 &         22 &         68 &      47223 \\
         2 &         23 &         56 &      49044 \\
         3 &         24 &         42 &      48586 \\
         4 &         25 &         38 &      47651 \\
         5 &         26 &         56 &      44973 \\
         6 &         27 &        139 &      49617 \\
         7 &         28 &         34 &      47905 \\
         8 &         29 &         35 &      47465 \\
         9 &         30 &         34 &      49603 \\
        10 &         31 &         23 &      49827 \\
        11 &         32 &         25 &      49406 \\
        12 &         33 &         63 &      45207 \\
        13 &         34 &         53 &      49803 \\
        14 &         35 &         26 &      45723 \\
        15 &         36 &         26 &      44137 \\
        16 &         37 &         30 &      44109 \\
        17 &         38 &         42 &      48621 \\
        18 &         39 &         38 &      49694 \\
        19 &         40 &         38 &      47725 \\
        20 &         41 &         80 &      44893 \\
        21 &         42 &         77 &      47874 \\
        22 &         43 &          8 &      23443 \\
\end{tabular}
\begin{tabular}{rrrr}
	\begin{sideways} Chromosome Number \end{sideways} & \begin{sideways} Region ID \end{sideways} & \begin{sideways} Number of SNPs \end{sideways} & \begin{sideways} Region Size (bp) \end{sideways} \\
	\hline
         1 &         44 &          6 &      44916 \\
         2 &         45 &         24 &      47963 \\
         3 &         46 &         16 &      46999 \\
         4 &         47 &         32 &      46961 \\
         5 &         48 &         31 &      49698 \\
         6 &         49 &         25 &      48619 \\
            &              &              &                  \\
            &              &              &                  \\
            &              &              &                  \\
            &              &              &                  \\
            &              &              &                  \\
            &              &              &                  \\
            &              &              &                  \\
            &              &              &                  \\
            &              &              &                  \\
            &              &              &                  \\
            &              &              &                  \\
            &              &              &                  \\
            &              &              &                  \\
            &              &              &                  \\
            &              &              &                  \\
            &              &              &                  \\
\end{tabular}
\end{table}

We then applied our new method to these data.  In all cases we set the
\Index{effective population size} for our model, $N_e$, to
15,000  and set $ \alpha = \alpha_1 = \alpha_2 $  (this may be
thought of as the simple situation when there is an equal
amount of gene flow in each direction).  

As a first test of
using the new model we decided to test the sensitivity to the
number of regions used and the number of SNPs selected across
those regions, where we specify the correct number of
populations in the sample ($K = 3$) and set $\alpha = 0.1$.  In
our experiments we ignore the information as to which
haplotypes come from which populations---this is what we aim
to infer, and we can then compare inferences with the truth. We denote by $r$ the number of independent regions chosen, where $r \in \{5,
10, 20, 30, 40, 50\}$, and by $s$ the number of SNPs chosen
in the selected regions, where $s \in \{10, 20, 50, 80, 100\}$.
For every pair of values $r$ and $s$ we independently at random
selected $r$ regions and then selected $s$ SNPs from the selected region to be included in the analysis.  We did this 20
times for each pair of values $r$ and $s$ and tested the method
on each of the resulting samples from our data.  

\begin{figure}
% \centering
  \includegraphics[width = \textwidth]{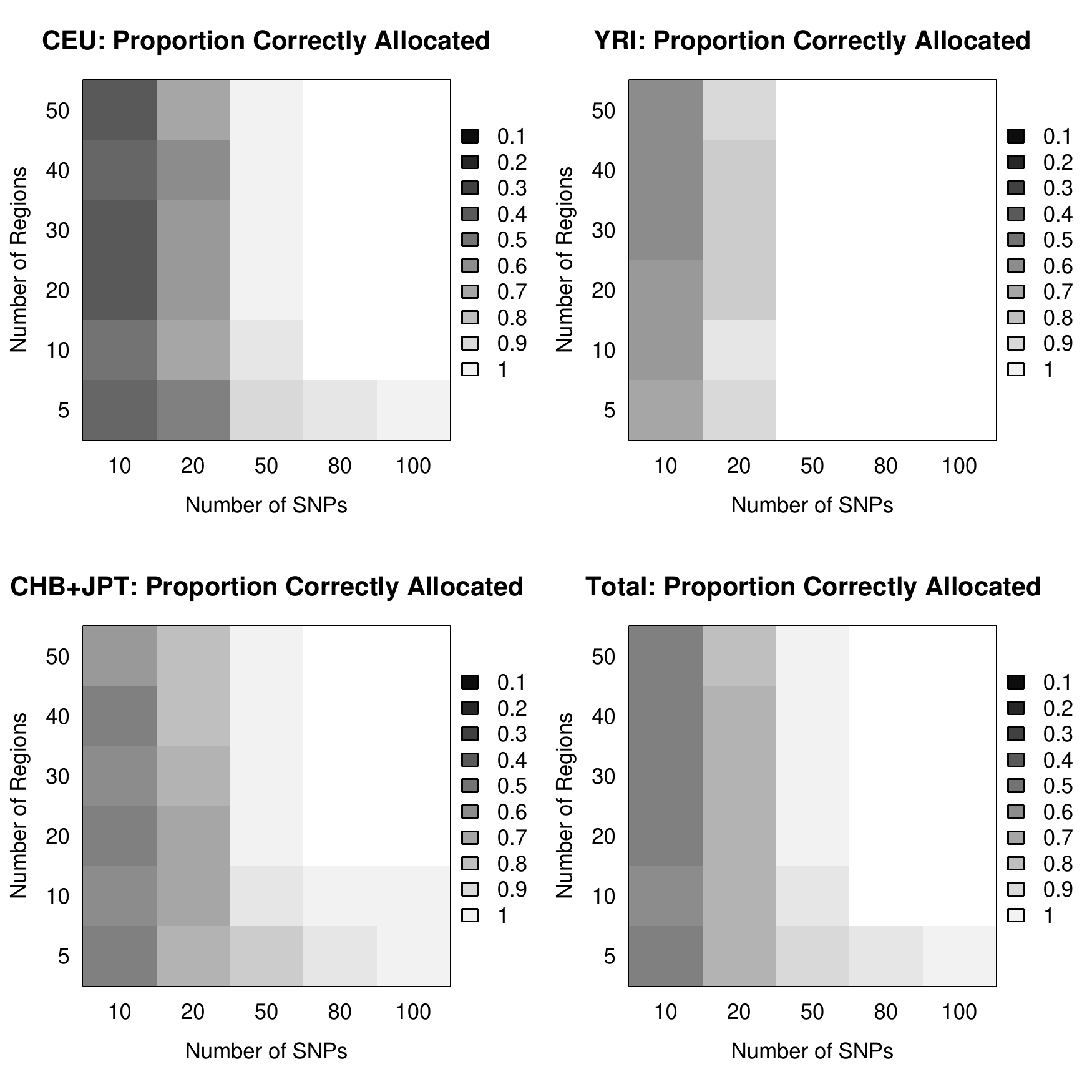}
  \caption{Proportion of haplotypes assigned to the correct population.  In
  this application of our method the number of populations is set to the
  correct number ($K = 3$), only SNPs with
\index{minor allele frequency (MAF)}minor allele frequency (MAF) $> 0.05$ are included in the analysis, and $\alpha = 0.1$.  Each entry at position $(r, s)$ in the four charts relates to a random selection of $r$ regions out of the 50 regions in the data, with $s$ SNPs then randomly selected for each region.  Values shown are averaged over 20 runs of the method in each case, with a burn-in of 200 iterations and 1000 samples kept after the burn-in.  Sequences are allocated to the cluster to which they are assigned for the majority of the iterations after the burn-in.  The top left chart shows the proportion of CEU haplotypes assigned correctly to the CEU cluster.  The second, third and fourth charts show the equivalent proportions for the YRI, CHB+JPT and Total Population respectively.}
 \label{fig:LD_resultsSNPsRandNoTrainMAF0.05}
\end{figure}

In each run we used a
\index{burn-in|(}burn-in of 200 iterations and then kept the following
1,000 iterations.  We ran several analyses to confirm that these values are
sufficient for convergence.  Haplotypes were assigned to the cluster in which
they spent the majority of iterations (after the burn-in).  To check whether
label-switching was occurring we kept the pairwise assignments matrix for each
run.  As with
\index{Pritchard, J. K.}\index{Stephens, M.}\cite{PSD2000} and
\index{Falush, D.}\cite{FSP2003}, label-switching was not
observed and thus the clusters are well-defined. In order to assess
performance, clusters were labelled by the population from which the majority
of their assigned samples were derived.  As we shall see, given the accuracy
of the method, this does not result in any ambiguity.  Performance was
measured as the proportion of haplotypes assigned to the correct population,
where we measured these proportions for each sample population individually,
and also for the combined sample.  We show the average performance over the 20
runs for each pair of values $r$ and $s$ in
Figure \ref{fig:LD_resultsSNPsRandNoTrainMAF0.05}.

%%%%
% NO TRAIN MAF 0.05 - FIGURE
%%%%

Examination of Figure \ref{fig:LD_resultsSNPsRandNoTrainMAF0.05} reveals some insights about the performance of the method.    As would be expected, across all populations, for each fixed value of the number of regions selected ($r$), average performance increases with the number of SNPs used.  Thus, the more SNP information we have, the more accurate are our population allocations.  In general, increasing the number of independent regions used has less effect on the accuracy of the population assignments, although accuracy does increase with increasing the number of regions used.   We conclude that
applying our method to data sets comprising at least 80 SNPs
derived from at least 10 independent regions will give high
accuracy ($\geq 95\%$) with good precision (the standard deviation over all runs observed in this case was less than $2\%$).

We then tested the effect of varying the $\alpha$ parameter, for $\alpha$ in
the range $(0, 0.3)$.  In this case we used a single set of SNPs for testing.
We used a set of 50 SNPs derived from each of 10 regions which had given
average accuracy over previous tests (approximately $90\%$ of individuals
classified correctly in the initial tests).  We selected this set as it
represents a typical possible application of the method and also gives
reasonable scope to examine the effect of $\alpha$ both in decreasing and
increasing accuracy.  For each value of $\alpha$ we ran the method 10 times
with randomly chosen initial assignments, using a burn-in of 200 iterations
and retaining 1,000 iterations after the burn-in.   We observe that for a
given value of               $\alpha$ up to 0.2 the \index{sampling|)}sampler
converges to virtually the same result over all ten runs for all population
groups.  For $\alpha$ in this range, performance varies across populations but
remains consistently above $90\%$ of total chromosomes\index{chromosome|)}
assigned correctly.  The best overall performance is observed when $\alpha =
0.01$.  For values of $\alpha$ greater than 0.2 the method no longer converges
and the average accuracy of assignments is reduced.  The number of individuals
classified correctly for various values of $\alpha$ in the range 0 to 0.2
differs by only a small amount, so it is difficult to draw too many
conclusions from the limited experiments we have performed.
\index{data!HapMap|)}Close examination of the posterior probabilities for all
individuals in the analyses may provide fruitful insights, although this is
left to future work.

In conclusion, the method performed well and, provided sufficient
\index{single nucleotide polymorphism (SNP)|)}SNPs
from sufficient independent regions are available for
the \index{haplotype|)}haplotypes that are to be classified, gives better than
95\% accuracy for classifying sequences into these continental populations.
The method converges rapidly: in most cases investigated, a
\index{burn-in|)}burn-in of 200 iterations with a further 1,000 iterations
retained for analysis was seen to be sufficient, but we would advocate the use
of more iterations than this.  It is feasible to apply the
method to data sets of the scale considered here.

Note that there are natural, more sophisticated, approaches to
handling the uncertainty associated with assignment of
individuals to populations.  Long-run proportions of time spent
in different assignments naturally lead to posterior
probabilities for assignment.  These could be carried through
into subsequent analyses, or thresholded at some (high) value,
so that population assignments are made only when these
posterior probabilities are close to
unity.\index{mathematical genetics|)}\index{population structure|)}

\paragraph{Acknowledgement}PD acknowledges support from the Royal Society.

\bibliographystyle{cambridgeauthordateCMG}
\bibliography{kingfest}

\end{document}